\documentclass[12pt.a4paper]{article}
\usepackage{graphicx}
\usepackage{amsmath}
\usepackage{amssymb}
\usepackage{amsbsy}
\usepackage{imakeidx}
\usepackage{setspace}
\usepackage{caption}
\usepackage{framed}
\usepackage{float}
\usepackage{lipsum}   
\makeindex
\date{}

\begin{document}
\setcounter{tocdepth}{5}
\setcounter{secnumdepth}{5}

\setcounter{page}{6}
\newpage
\pagenumbering{arabic}
\setcounter{page}{1}
\large
\baselineskip25pt
\title{Effect of Meditation on Heart Rate Variability}

\author{\\[0.4in]Sobhendu Kumar Ghatak \\ Department of  Physics, Vidyamandir, Belur, 700123, India }

\maketitle
\newcommand{\be}{\begin{equation}}
\newcommand{\ee}{\end{equation}}
\newcommand{\ba}{\begin{eqnarray}}
\newcommand{\ea}{\end{eqnarray}}
\def\Eins{{\mathchoice {\rm 1\mskip-4mu l} {\rm 1\mskip-4mu l}
           {\rm 1\mskip-4.5mu l} {\rm 1\mskip-5mu l}}}

\begin{abstract}
The dynamics of the autonomic nervous system,a prime factor for variation of heart rate is modulated by life style factors like meditation and breathing excercise.The effect of meditation ('\textit{dhayan}') practiced by disciples of \emph{Ramakrishana Mission}(RKM) on the heart rate variability is examined. The heart rate (HR)of a group of disciples of RKM is obtained from ECG signal recorded over short time duration. Lagged Poincar\'{e} plot of the heart rate(HR),the method of the principal component analysis and the autocorrelation of HR fluctuation are used to analyze the HR.The variation of Poincar\'{e} parameters $(SD1)$,$(SD2)$ and their ratio $(SD12)$, with lagged number are non-linear,and reveal a significant change after meditation. In particular,The magnitude of the slope and the curvature of $(SD12)$  with lagged number increases after meditation.The meditation reduces the heart rate and increases the stroke volume.A strong correlation between slope and curvature of $(SD12)$ is noted from the correlation matrix of multi-dimensional data set resulting from the Poincar\'{e} plot.The principal components (PC) of data resulting from pre- and post- meditation are well separated in PC-space.The entropy associated with $R$-wave fluctuations for all meditators is reduced after meditation.The auto-correlation of HR fluctuation exhibits highly correlated pattern after meditation.The study suggests that the meditation improves the heart rate dynamics and the 'calmness' of mind.
\end{abstract}
\vspace{0.2cm}
\textit{e-mail for correspondence} :skghatak@phy.iitkgp.ac.in

\newpage
\section{\textit{Introduction}}
Heart rate variability (HRV),a measure of fluctuations of heart rate (HR) obtained from the measurement of two consecutive R-R interval in ECG signals is of importance in assessing health of living being \cite {task} -\cite {Acharya}.The autonomous nervous system(ANS)is prime factor for modulating HR,and thereby any change in ANS is reflected in HRV.The measurement of heart rate variability(HRV) which is  non-invasive,sensitive,faster and reproducible is successfully utilized to find the influence of any factor that alter cardiovascular autonomic function.A number of indices of HRV obtained from the detailed analysis seem to establish notable relationship between ANS and various diseases \cite {Eberhard}\cite{Sztajzel}.The sympathetic and parasympathetic branches of ANS affects cardiac rhythm in non-additive fashion and any change in these branches results corresponding changes of the parameters of HRV. \\

The HRV is quantified from HR data using linear measures in the time and frequency domain.The data acquired for long time provides better measure in frequency domain. However,long time ECG recording is inconvenient for subjects staying in a given position and also time consuming.The spectral analysis in frequency assumes stationarity of signal and sudden changes in HR results alteration of power spectrum which often difficult to interpret.The power spectrum of HR signal provides the power in high frequency (HF) band ($\approx 0.15–0.4$ Hz) representing mostly of parasympathetic response and low frequency (LF) band ($ \approx 0.04-0.15$ Hz) resulting from combined influence of the sympathetic and that (albeit small)the parasympathetic responses.The ratio of power LF/HF is often considered as a measure of sympathovagal balance \cite {task}.Simpler method to assess complex non-linear behavior in the study of physiological signals is Poincar\'{e} plot \cite{Eberhard}, \cite{Kamen},\cite{Piskorski} .
The Poincar\'{e} plot,that reflects non-linear aspect of cardio-dynamics,depicts graphically HR fluctuation and it is scattered plot where each RR interval is plotted against its next interval.
The plot uses unfiltered data and simpler to study dynamics of heart rate variability.The scattered plot of RR intervals forms a cluster and the visual inspection of the shapes of the cluster in the Poincar\'{e} plot is a guide to assess the quality of recorded ECG signals and identification of premature and ectopic beats \cite{Mourot},\cite{Sosnowski}.The studies based on short-term HRV with epochs as short as $300$ beat to beat interval(around $5$ minutes of ECG recording) in time domain suggest that the measures are better reproducible than frequency domain measures \cite {Sztajzel}.The measures are different in two domains but are not mutually independent due to strong correlation between the parameters.As a result any additional information useful for better discrimination of subjects with cardiovascular deregulation are limited \cite {Liao}.Therefore,the search for new parameters,which are able to provide additional information embedded in the HRV signals is pertinet.In the plot, it is implicitly assumed that two successive R-R intervals are well correlated.This assumption lends itself to further generalization to lagged Poincar\'{e} plots by plotting $RR_{i+m}$ against $RR_{i}$ where $m$ represents the distance (in number of beats) between beats.Due to nonlinear nature of heart dynamics, a nonlinear analysis of HRV would provide more details of the dynamic process \cite{Eberhard}, \cite{Bergfeldt},\cite{Voss},\cite{Bhaskar},\cite{Sobhendu}.The Lagged Poincar\'{e} method,an extension of conventional Poincar\'{e} one has been found to be a better quantitative tool \cite {Karmakar} and is substantiated from some studies on Chronic Renal failure (CRF) \cite{Lerma},Congestive Heart Failure (CHF)\cite{Thakre}, diabetic subjects \cite{Contreras},\cite{Bhaskar} and rotatory audio stimulation \cite{Bhaskar Roy}.Different methodology for non-linear analysis of HRV are being examined \cite{Voss}. The methods like 3-D return mapping \cite{Ruy}, wavelet analysis \cite{Acharya},Detrended fluctuation analysis (DFA)\cite{Zhi}, \cite{Hu}, \cite{{Eduardo}} have been applied to extract different aspects of variability of heart rate.The principal components analysis of Poincar\'{e} parameters  with hypertension,diabetic and control group have pointed importance of cardiovascular dynamics \cite{Sobhendu}.\\
The meditation,a practice that augments mind-body relationship often used in life-style-medicine\cite{Dean}.There are different types of meditation ,e.g Mindfulness,Healthfulness Vipasayana and \textit{Dhayan}.The effect on the heart rate variability due to meditation procedure has been studied in time domain \cite{Narendra}\cite{Anne}\cite{Chu}\cite{Takahashia} with different. In this work,the heart rate variability of twenty disciples of \textit{Ramkrishna Mission} who practice \textit{Dhayan} regularly,are studied in time domain from ECG signal recorded before and after performance of meditation.The HR variability is mainly analyzed using measures obtained from HR analysis in time domain.The results of the lagged behaviour of the Poincar\'{e} parameters,the Principal component analysis (PCA),the entropy associated with $R$ wave and the correlation of HR fluctuation are presented.The study is aimed to explore and assess health benefit of meditation(\textit{Dhayan}).

\section{\textit{Method}}
The disciples (\textit{swami})from \textit{Ramkrishna Mission,Belur} twenty in numbers agreed to volunteer for this endeavour. They are regular meditator and with good health. Each of them was explained about the object of study,and non-invasive aspect of ECG.The written consent from each participant was taken. Except two senior disciples,
the age of other disciples are in the range of $24$ to $36$ years.The ECG data were recorded in supine position for $6$ minutes before and after \textit{Meditation(\textsl{dhayan})} at \textit{mission campus}.The sampling rate of ECG signal with II-lead  configuration $500$Hz. The RR interval were extracted from the ECG data with the help of ORIGIN software.The peaks other than regular one were discarded in the analysis.\\
\subsection{\textit{Poincar\'{e} Plot}}

A scatter plot of duplets of successive $R-R$ interval ($RR$) is the Poincar\'{e} plot. Further generalization of Poincar\'{e} plots is obtained by plotting m-lagged plots where m represents the distance(in number of beats) between the duplet beats, that is, the'lag' of the second beat from the first one \cite {task},\cite {Sztajzel},\cite{Bergfeldt},\cite{Kamen},\cite {Karmakar}. Three parameters namely $(SD1)$,$(SD2)$ and $SD12 = SD1/SD2$ quantify the character of the plot.The parameter $(SD1)$ is the measure of the standard deviation of instantaneous variability of beat-to-beat interval  and $(SD2)$ is that of the continuous long-term RR interval variability.The relative measure $SD12 = SD1/SD2$ provides non-linear aspects $RR$ interval.Any activity that stimulates ANS such as meditation causes the change in peak value of $R$.Such effects can be examined with the help of similar plot successive peak value($Rpk$) and a significant change has been found in this study.

\subsection{\textit{Principal Components Analysis}}
Principal component analysis (PCA) is a statistical procedure by which a large set of correlated variables can be transformed to a smaller number of independent new set of variable without throwing out essence of original data set\cite{Jolliffe}.The new set of variables are referred as the principal components (PCs).They are linear combination of original one with weights that form orthogonal basis vectors.The diagonalisation of
the covariant matrix of data set gives these basis vectors-e.g. eigenvectors and eigenvalues with decreasing value.The
 Each PC's are linear combination of  data weighted with the eigenvectors and carries new information
about the data set. The first
few components of PC normally provide most of the variability of data set.
In this study, two sets of data are considered for the PCA analysis.First set contains the parameters obtained from extended Poincar\'{e} plot \cite {Jolliffe},\cite {Mika}.The variables namely $SD1$,$SD2$,$SD12$ for lag $m=1$ and the maximum value of the slope and the curvature of these variables with lag $m$ are considered assuming that these parameters specify important features of the HRV \cite{Bhaskar Roy},\cite{Sobhendu}. The parameters are in different units and are normalized using the norm of respective parameter of the group.The normalized and mean subtracted data sets are presented as matrix $X_{M,N}$ where $M$ rows and $N$ columns refer to number of subjects and parameters respectively.Second set of data contains five time-domain parameters namely -$HR$ ,its fluctuation $STD$,the magnitude of R-wave $Rpk$, its $STD$ and the entropy $S$ obtained from the fluctuation of $Rpk$ values.

The covariance matrix $C$ of normalized data was then obtained as
\begin{equation}
  C = \frac{1}{M - 1}X^{T}X
  \label{matrix}
\end{equation}

The eigenvalues and the eigenvectors of matrix $C$ are obtained for each group using MATLAB programme.Out of $N$ eigenvalues it turns out that the first $n$ (according to their magnitudes)values can account most of the data and the corresponding $n$ eigenvectors are considered as basis vectors for calculation of principal components.The eigenvector is given by

\begin{equation}
 \Psi_{N,n} =
 \begin{pmatrix}
  \varphi_{1,1} & \varphi_{1,2} & \cdots & \varphi_{1,n}\\
  \varphi_{2,1} & \varphi_{2,2} & \cdots & \varphi_{2,n}\\
  \vdots  & \vdots  & \ddots & \vdots \\
  \varphi_{N,1} & \varphi_{N,2} & \cdots & \varphi_{N,n}
   \end{pmatrix}
 \label{eigvec}
\end{equation}
The principal components $PC$ can then be calculated from the  matrix equation
\begin{equation}
  PC_{n,M} = \Psi^{T}X^{T}
  \label{pc}
\end{equation}
where  $PC$ is the matrix of {nxM}.The eigenvector is the measure of the weight of data in determining the principal components and so $PC_{n,M}$ is $n$ principal component of $M$ subject.
The elements of covariant matrix represent the amplitude of correlation between the Poincar\'{e} parameters.
\subsection{\textit{Correlation Function}}

It is observed that there are significant changes in heart rate and the peak value of $R$-wave,and the nature of changes are examined through correlation of both heart rhythm and $R$-peak before and after meditation.The time-series of the increments of ($RR_{i}$) was constructed and the correlation of deviation of successive beat interval was calculated.
The autocorrelation function $Corr-RR$ was obtained from the equation
\begin{equation}
Corr-RR(m) = \frac{1}{\sigma} \sum_{i}^{N}\Delta RR_{i}\Delta RR_{i-m}
\label{cor}
\end{equation}
where $\sigma$ is the normalizing factor ($Corr-RR(0)$  with zero lag $m$) and $\Delta RR_{i} = RR_{i+1} - RR_{i}$ is the deviation between two consecutive beat interval.
Similarly, the correlation function $Corr-Rpk$ of amplitude of peak value of $R$ wave was obtained replacing $RR$ by $Rpk$ in above equation.

\section{\textit{Results and Discussions}}
\subsection{\textit{The Poincar\'{e} Parameters}}

A representative scattered Poincar\'{e} plot of $RR$ interval and peak value of $R$ signal $Rpk$ for a subject is shown in Fig.$1$. The left part of figure is for $RR_{i}$ - $RR_{i+1}$ and that on the right for $Rpk_{i}$ - $Rpk_{i+1}$ ,and the corresponding data points, marked by red-dot and green-dot, are for pre- and post- meditation.

\begin{figure}[h!]
\centering
\includegraphics[width=1.1\textwidth]{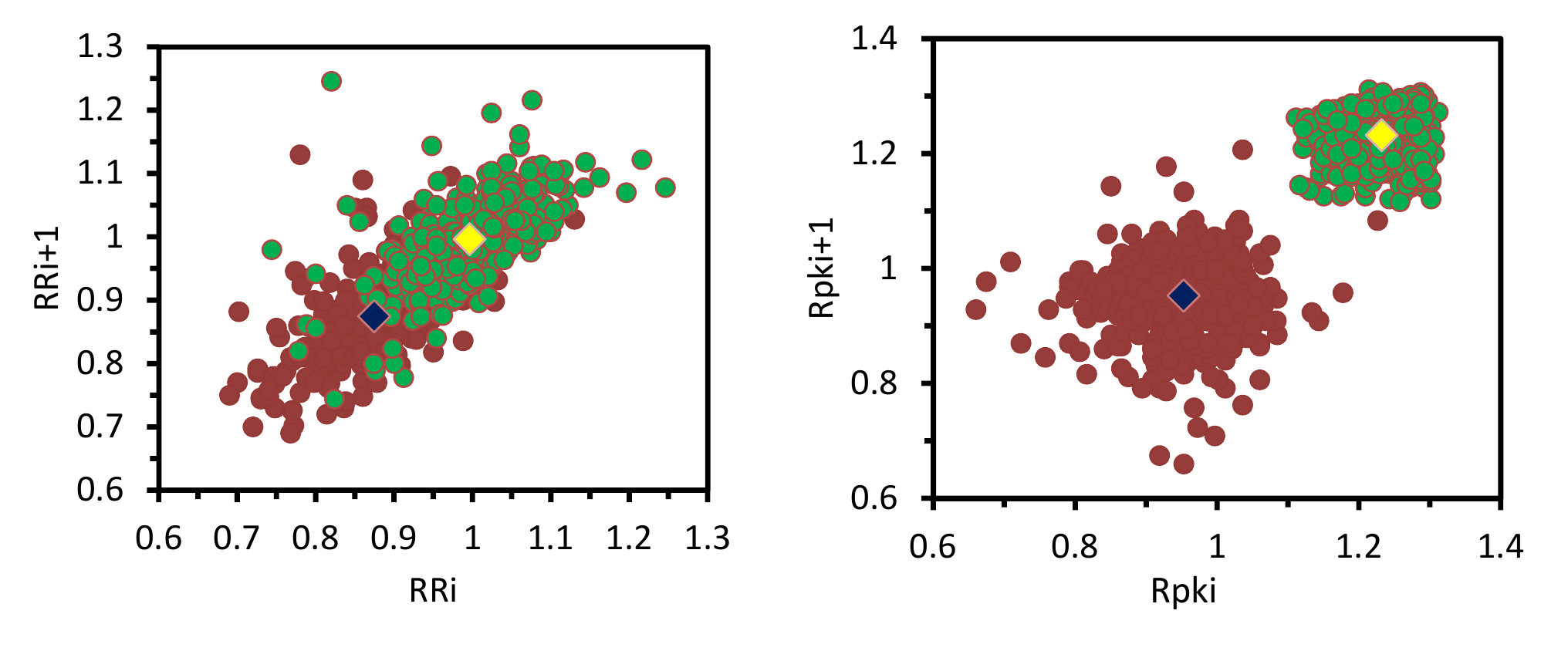}
\caption{The Poincar\'{e} plot of $RR_{i}$ and $RR_{i+1}$ (a), and similar plot of amplitude $Rpk$ of $R$ wave (b) before (Red-dot) and after (Green-dot) meditation.The blue and yellow diamond are respective mean values.Both results are for a young monk.}
\label{pcareplot}
\end{figure}
The depiction shows the $RR_{i}$ interval after meditation becomes longer and so heart rate is reduced. The mean values  of $RR$ intervals before (blue-diamond) and after (yellow-diamond) meditation are well separated.The plot of the magnitude $Rpk$ of $R$-peak increases due to meditation and more importantly, the scattering of data is much reduced (Fig.\ref{pcareplot}).As $Rpk$ is associated with pumping action, and so it indicates a better ventricular activity of heart after meditation.This general trend was observed for all subjects though amount of scattering and shift of mean values differ from person to person.

The Poincar\'{e} indices help to analyze real-time analysis of short duration of ECG signal.
The values of $(SD1)$ and $(SD2)$ were calculated for lag$ = m$ from the relations
$SD1(m) = (\Phi(m) - \Phi(0))^{1/2}$ and $SD2(m) = (\Phi(m) + \Phi(0))^{1/2}$, where the auto-covariance function $\Phi(m)$ is given by
$\Phi(m) = E[(RR_{i} - RR_{M}) (RR_{i+m} - RR_{M})]$
and $RR_{M}$ is the mean of $RR_{i}$ \cite{Brennan},\cite{Bhaskar},\cite{Ghatak}.
We first analyze the results of indices for $m=1$.

\begin{figure}[h!]
\centering
\includegraphics[width=0.8\textwidth]{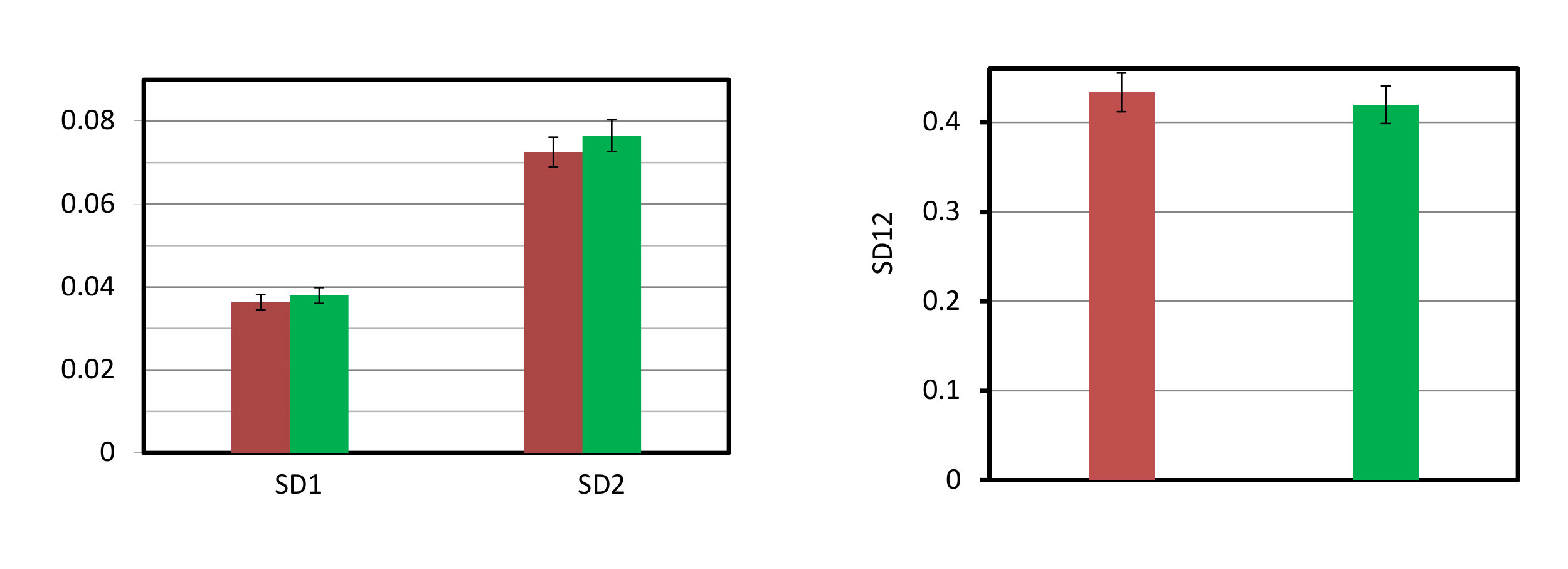}
\caption{The box plot of group mean value of $(SD1)$,$(SD2)$ and $SD12$ before (Red)and after (Green) meditation for lag $m=1$.All changes are statistically significant at level ($p < 0.05$).}
\label{sdbox}
\end{figure}

The group mean values of $SD1$,$SD2$ and $SD12$  for $m=1$ are presented in Fig.\ref{sdbox}.Both short-term ($SD1$) and long-term ($SD2$) variance of HR are higher but the ratio ($SD12$) is lowered after meditation.The analysis of an extended Poincare\^{e} plot points out importance of $SD12$ and opens up for new measure of HRV.The mean values of $SD1$,$SD2$ and $SD12$ for different values of $m$ are obtained from $RR$ interval of individual subject before and after meditation.\\

The mean values of the parameters for all participants are an increasing function of lag variable $m$. The growth rate of these parameters are highest close to $m=1$ and decreases with $m$. This non-linear variation can be characterized by the slope and the curvature of plot.It is found both the slope and the curvature decrease with $m$.For large value of $m$ the curvature becomes negligible as $SD12$ approaches to its saturation.To estimate the slope and the curvature we used the Pad\'{e} approximation \cite{Bhaskar}.A simple form for the Pad\'{e} approximation is chosen for analysis of non-linear variation of these parameters
\begin{equation}
\centering
 Y = \frac{a+bm}{1+cm}
 \label{pade}
\end{equation}
where $Y$ is either $SD1$,$SD2$ or $SD12$ and is represented by the ratio of linear polynomial of $m$ with three adjustable parameters $a$,$b$ and $c$.\\
\begin{figure}[h]
\centering
\begin{minipage}[c]{.5\textwidth}
\begin{center}
\includegraphics[width=1\linewidth]{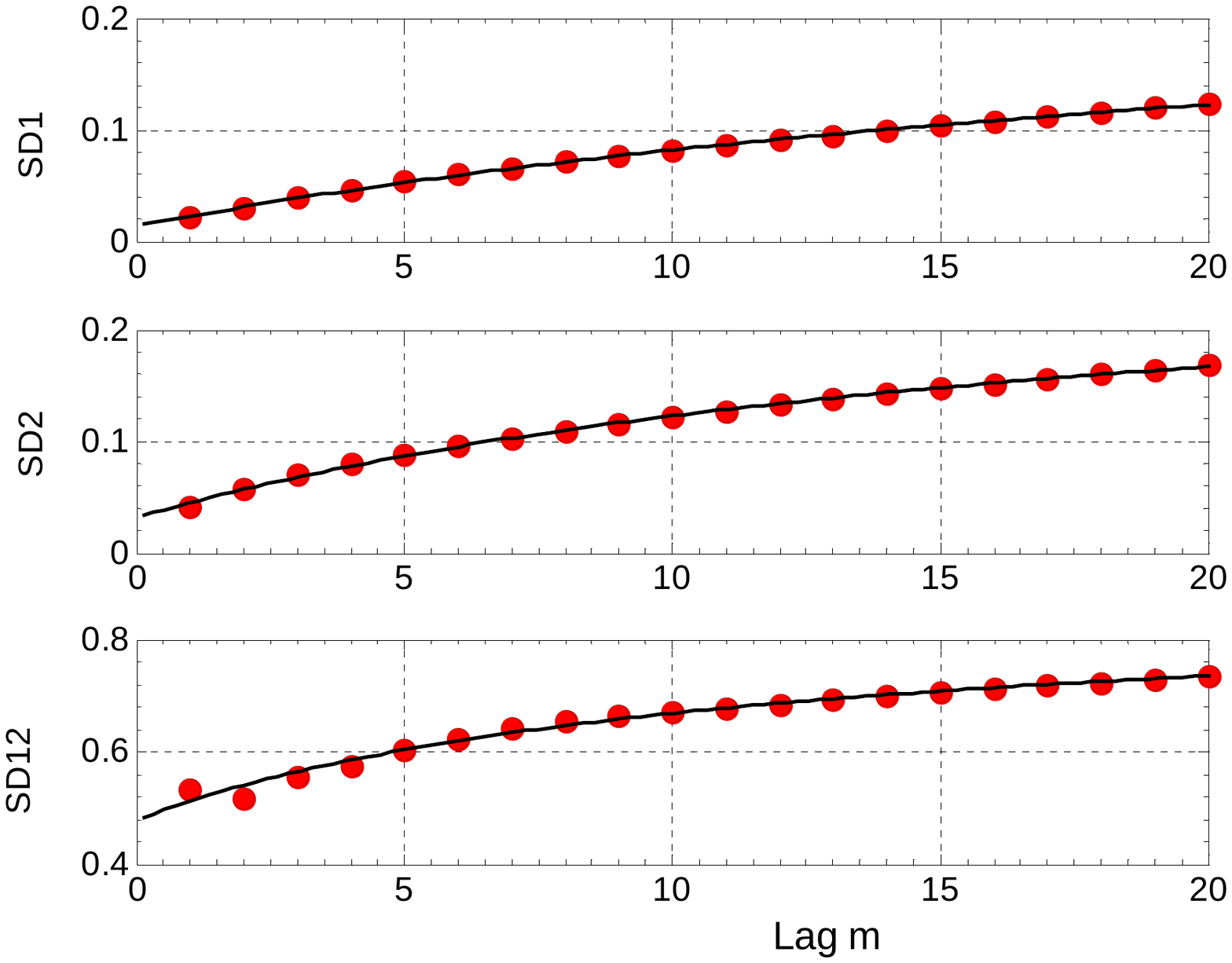}
  \end{center}
\end{minipage}\hfill
\begin{minipage}[c]{0.5\textwidth}
\begin{center}
\includegraphics[width=1\linewidth]{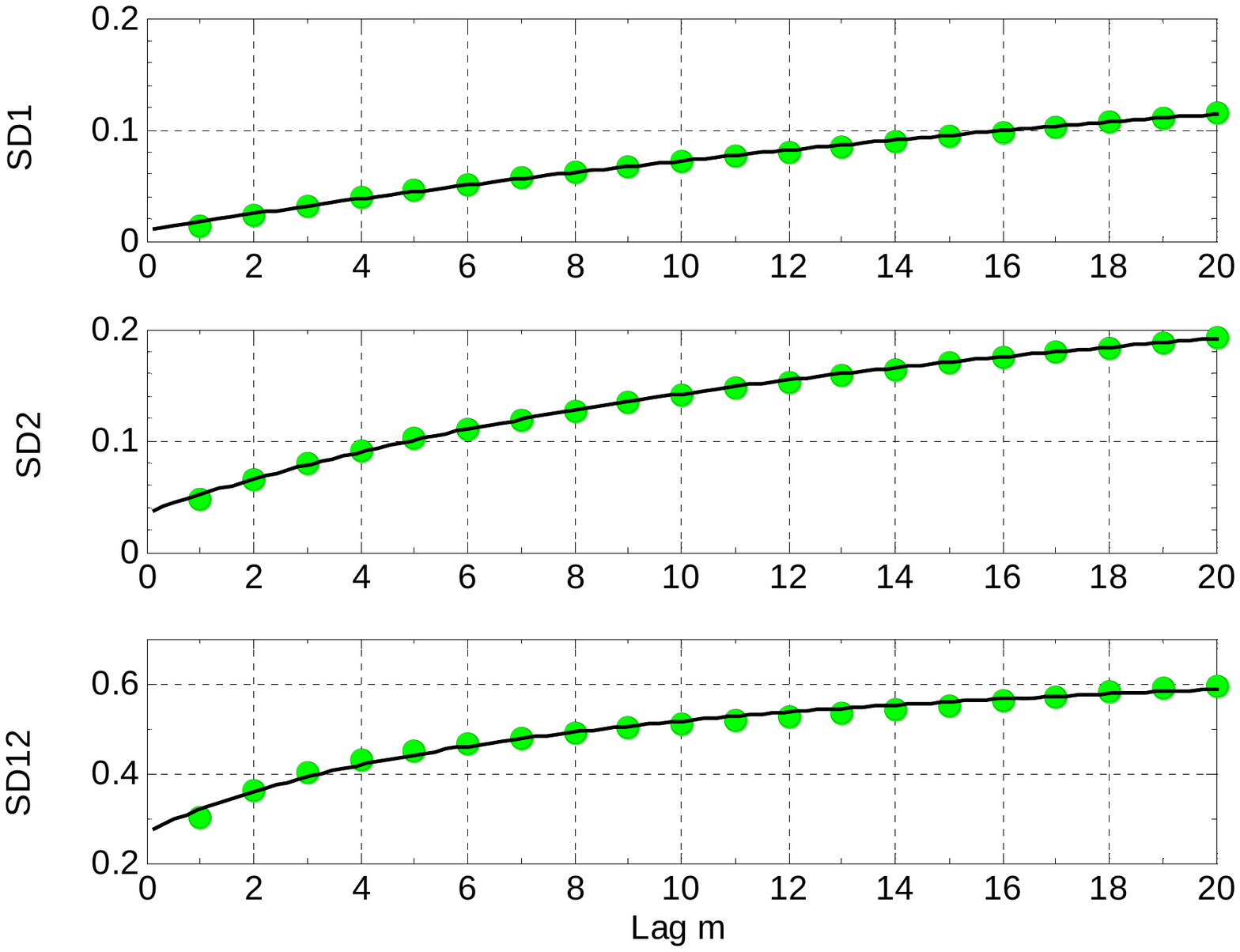}

\end{center}
\end{minipage}
\caption{Plot of the group mean values of $SD1$(upper),$SD2$(middle) and $SD12$(lower)and fitted curve(line).The left-figure
is for Pre- and right-one for Post- meditation period data. Fitted line is obtained with $R^2\simeq 99.98$ percent for all}

\label{SDSanjib}
\end{figure}

The data for each participant are fitted using non-linear fit with MATLAB programme.The parameters are obtained from the best fitted curve with $R^2 \simeq 99.98$ percent,and $ \chi^2 \simeq 10^{-6} $. A representative figure with data and the fitted curve is displayed in Fig.(\ref{SDSanjib}) .The slope and curvature of fitted curve had their large magnitude near $m=1$.

\begin{figure}[h!]
\centering
\includegraphics[width=0.9\textwidth]{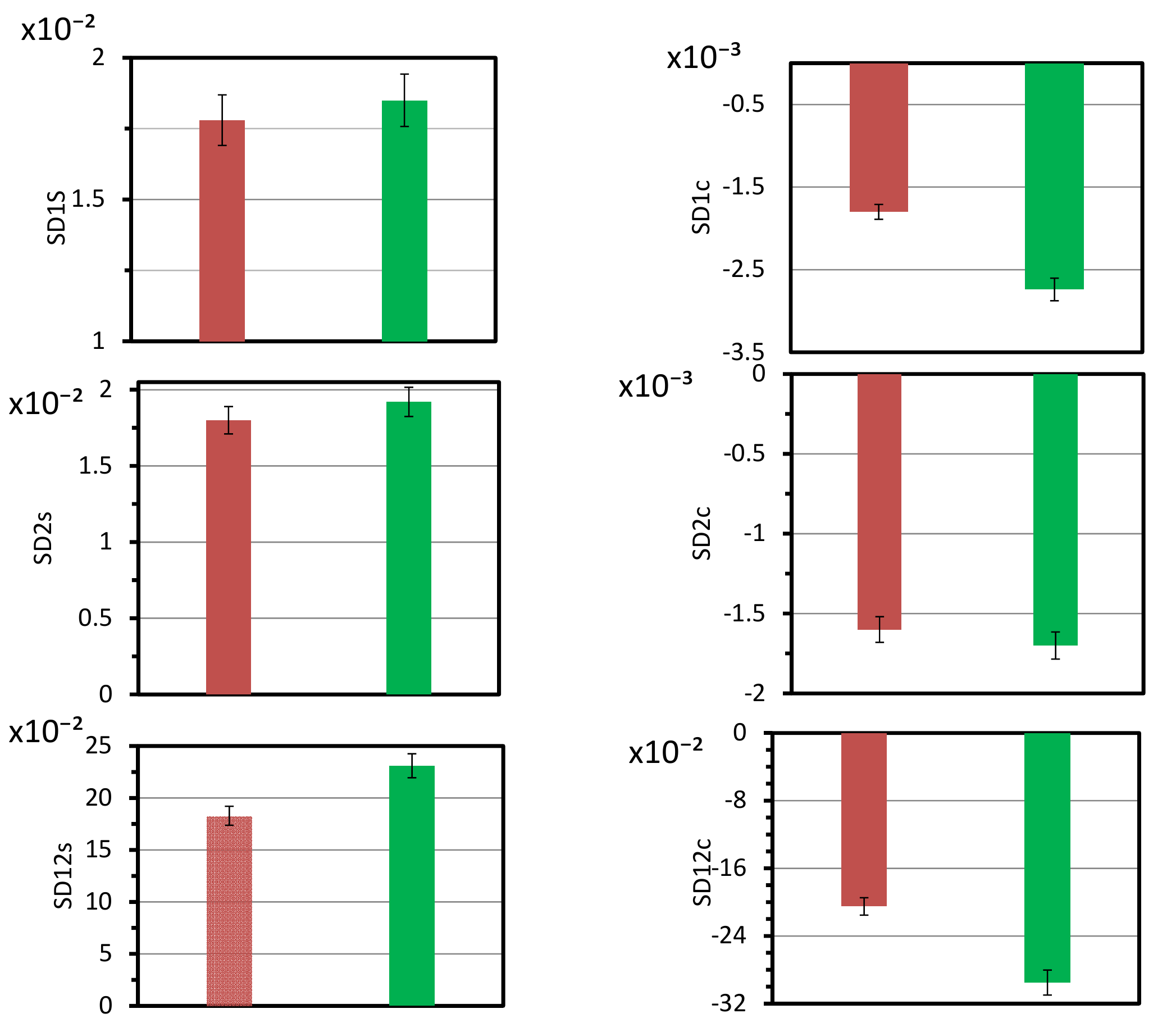}
\caption{The columns represent the group mean value of maximum of the slope and the curvature of respective plots of $SD1$ ,$SD2$ and $SD12$.The maximum value of these parameters appears close to lag $m =1$. The red and green refers to situation before and after meditation. Note the scale difference}
\label{sdall-box}
\end{figure}

The maximum values of the slopes $SD12s$ and the curvature $SD12c$ of $SD12 -m$ plot for individual subject within each group are considered.The results of group mean are summarized in Fig.\ref{sdall-box}.The magnitudes of both slopes and curvature are found to be higher after meditation.The lower magnitude of the curvature of $SD12$ was associated with the deviation of normal cardiovascular function of patient \cite{Thakre}.It is expected that this curvature tends to zero for non-innervated heart and is large for heart of healthy person whose heart beats are less constrained.

\subsection{\textit{ Principal Component Analysis of Poincar\'{e} Parameters}}

The parameters $SD1$,$SD2$,$SD12$ and their growth with lag $m$ namely-the maximum values of the slopes and the curvature  before and after meditation describe most aspects of the Poincar\'{e} plot.The principal component analysis of these data is used to envisage the relative importance of these $9$ parameters.
The covariant matrix has been constructed with the normalized values following Eq.(\ref{matrix}) and the matrix element represents the correlation coefficients.
\begin{figure}[h!]
\centering
\includegraphics[width=1.0\textwidth]{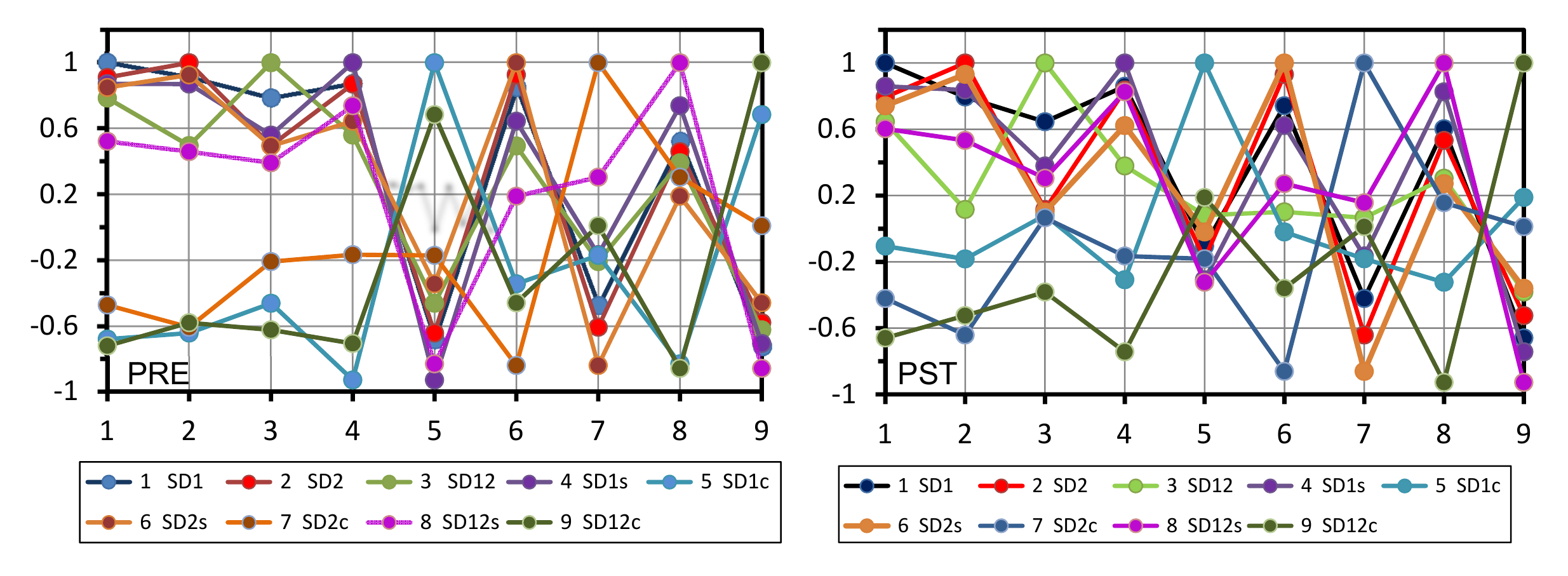}
\caption{ Plot of the correlation amplitudes which are elements of covariant matrix of before (PRE) and after (POST) meditation.The x-axis number represents the SD-parameters,marked ($1-9$) as indicated below the figure and y-axis the magnitude of components of correlation matrix}
\label{corr}
\end{figure}
The correlation matrix of Poincar\'{e}parameters derived from normalized data of Pre- and Post- are displayed in Fig.\ref{corr}.The slope and the curvature of a particular Poincar\'{e} parameter are highly correlated and negative value of the correlation coefficient is due to opposite sign of the slope and the curvature.The parameters $SD1$,$SD2$ are strongly correlated.On the other hand,$SD12$ is less dependent on other parameters.The slope and the curvature of $SD12$  are strongly correlated.The magnitude of the correlation aspects are augmented due to meditation.For example,the correlation coefficient between slope and curvature of $SD12$ is increased from $0.85$  to $0.93$.Such correlation between slope and curvature of $SD1$ is reduced in meditative state whereas that of $SD2$ shows opposite trend.

\begin{figure}[h!]
\centering
\includegraphics[width=0.8\textwidth]{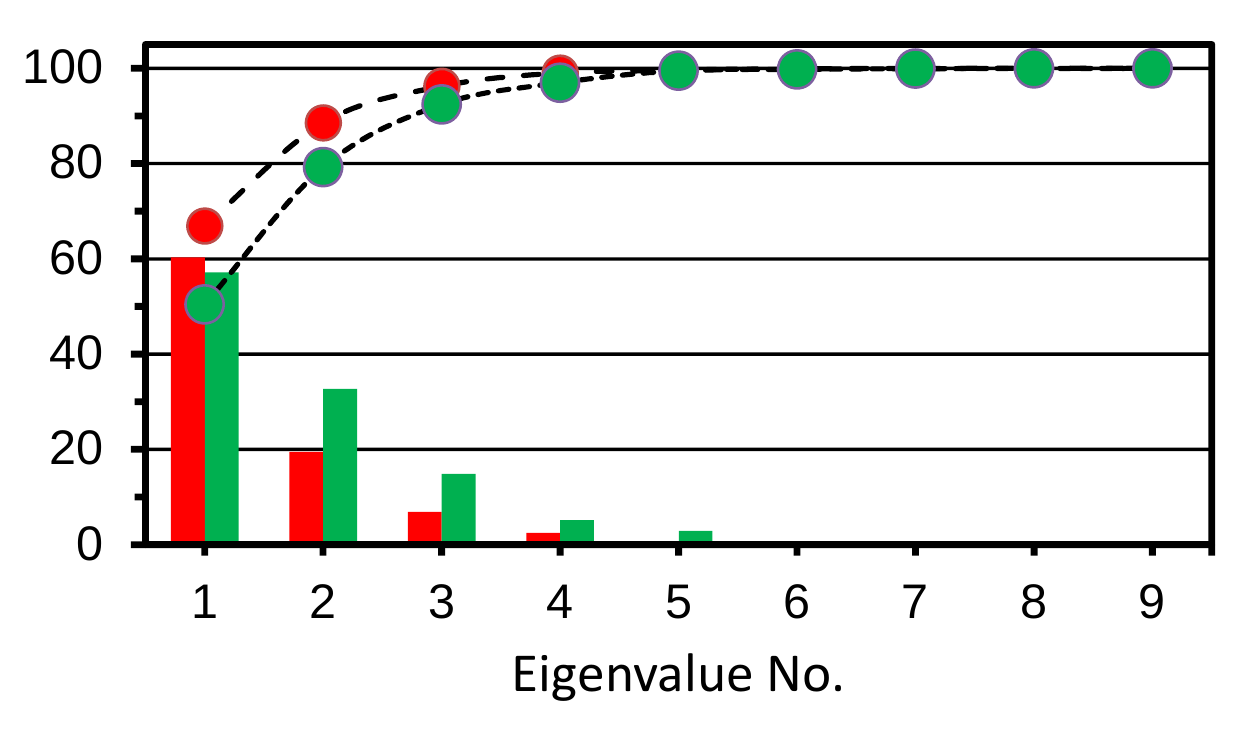}
\caption{Plot of Eigenvalue(column)and weight(dotted line) of each eigenvalue with Eigenvalue number.For presenting in same scale eigenvalues are multiplied by thousand.Red and green correspond to before and after meditation.}
\label{egn}
\end{figure}
There are nine eigenvalues of the correlation matrix. The eigenvalue decreases rapidly with eigenvalue number indicating that only few are relevant (Fig.\ref{egn}).The weight(dotted line) of eigenvalue indicates relative importance of eigenvalue in determining principle components.The value $100$ refers to complete data retrieval.It is evident that only first four eigenvalues account $99$ and $97$ percents of Poincar\'{e} parameters of Pre- and Post- situation.\\

\begin{figure}[h!]
\centering
\includegraphics[width=1.0\textwidth]{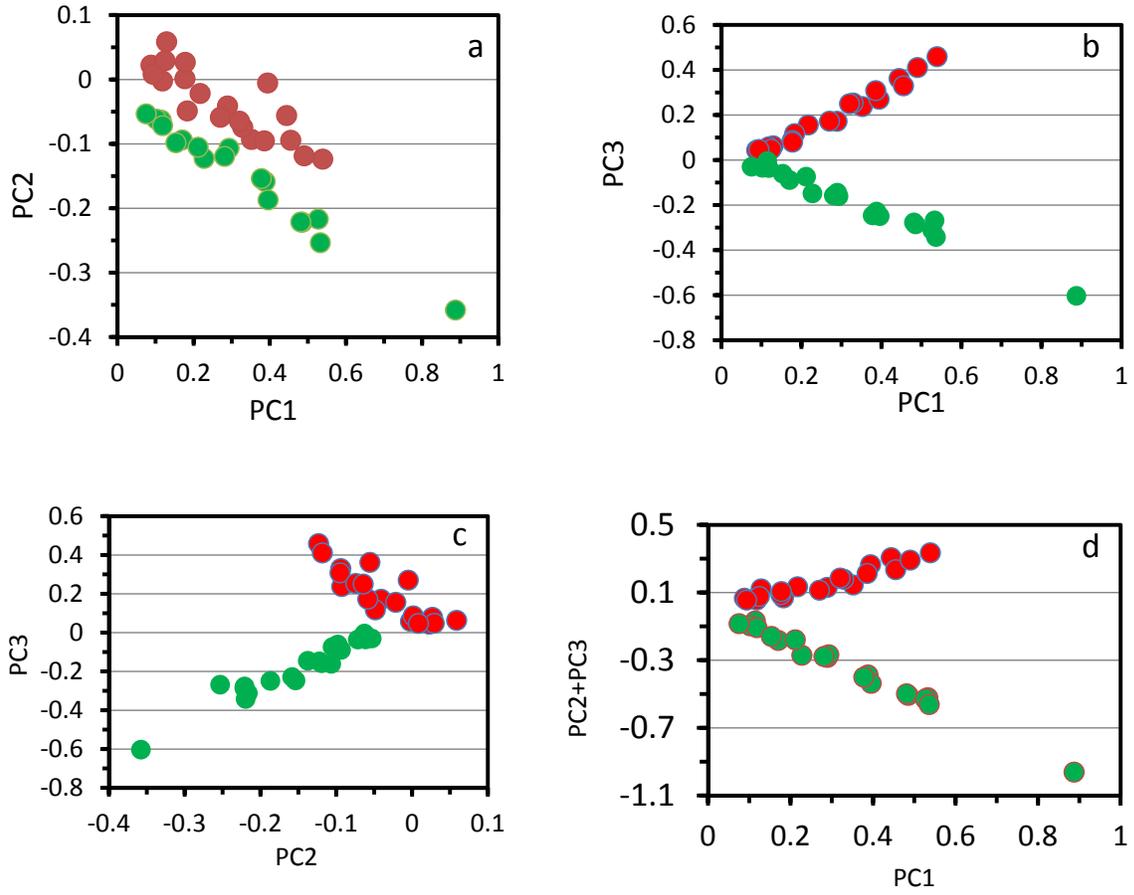}
\caption{Plot of values of $PC$ follows from normalized $9$ Poincar\'{e} parameters. Figs. a,b,d are results of $PC2$,$PC3$  and $PC2+PC3$ against $PC1$ whereas c for the variation of $PC3$ with $PC2$.The data marked with Red and Green are respectively before and after meditation. }
\label{pcsd}
\end{figure}
The principal components corresponding to four significant eigenvalues of individual person for Pre- and Post- meditation are obtained using eq.(\ref{pc}).We analyse the result with respect to the first principal component $PC1$ that corresponds to highest eigenvalue.The variation of other components (either single or combination) vis-vis first one are then expected to elucidate importance of spectrum of PCA.The trend of other components like second $PC2$, third $PC3$ and the sum $PC2+PC3$ visa-vis $PC1$ is displayed in Fig.\ref{pcsd} (\textit{a}),(\textit{b}) and (\textit{d})respectively.The Fig.\ref{pcsd}(\textit{c}) depicts the variation of $PC3$ with $PC2$.First thing to note is the concentration of the data into different region in phase space of the principal components.The data before and after meditation are well separated (Fig.\ref{pcsd}). For each participant there are significant changes of $PC$ indicating that HRV are altered in meditative state.

\begin{table}[h!]
\centering
\includegraphics[width=1.0\textwidth]{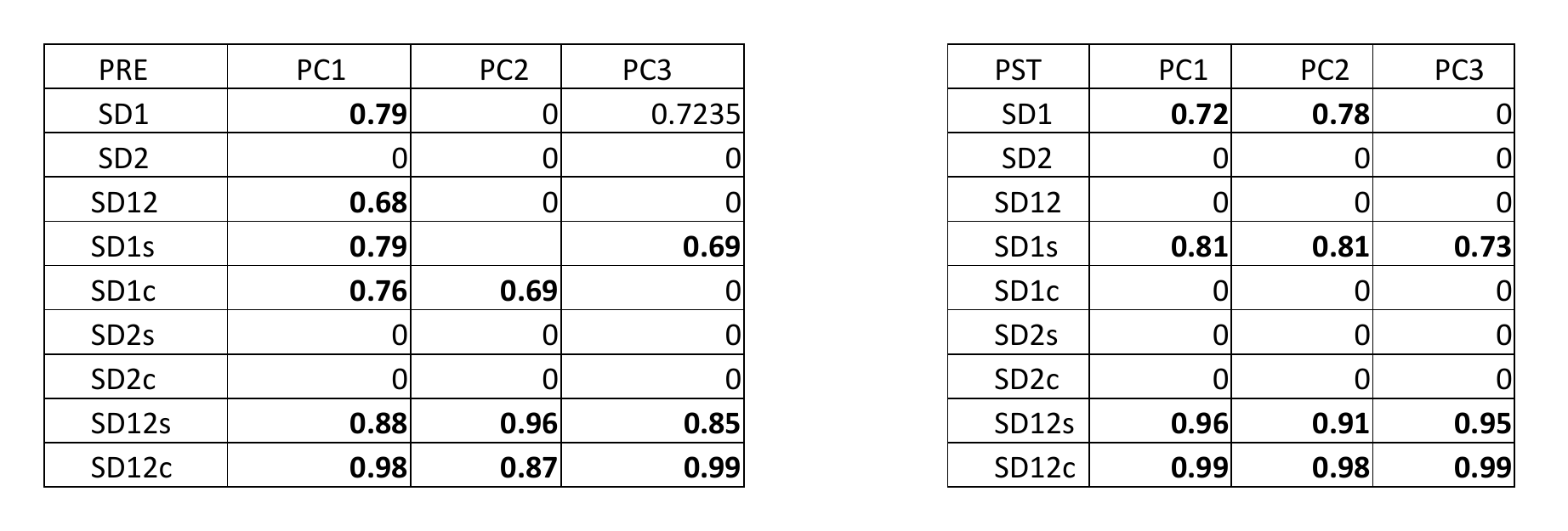}
\caption{The correlation amplitudes among $PC$ and the $9$ derived variables listed in rows. }
\label{cor-pc-sd}
\end{table}
 The correlation strength between $PC$ and nine derived variables are calculated along with $p$ values. The correlation coefficient with $p<0.005$ are tabulated given in Table.\ref{cor-pc-sd}.The zero values refer to case where $p$ values are larger than above limit.The only $6$ variables contribute to $PC1$.The short-time $RR$ interval fluctuation and its time evaluation are found to be more important than long-time one and their ratio.However,the growth of $SD12$ measured by maximum value of the slope and the curvature played dominant role in all three $PC$ and their contributions are augmented after meditation.

\subsection{\textit{ Principal Component Analysis of time domain parameters}}
The parameters that exhibit changes due to meditation are the heart rate $HR$,amplitude $Rpk$ of $R$ wave and their standard deviations $STD$ and the entropy $ENT$ obtained from fluctuation of $Rpk$.
\begin{figure}[h!]
\centering
\includegraphics[width=0.9\textwidth]{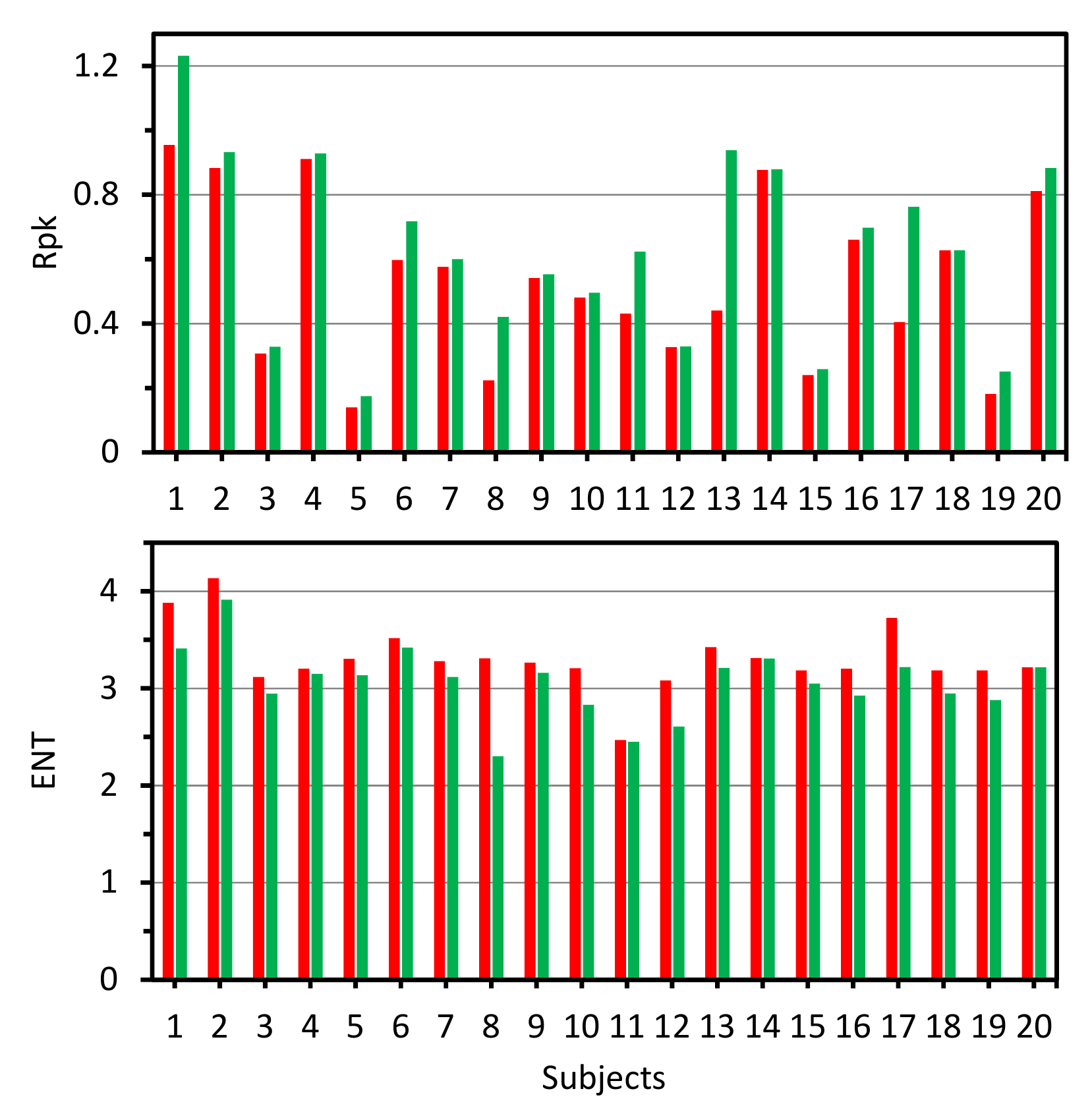}
\caption{Columnar plot of amplitude of peak $Rpk$ of $R$ (upper) and entropy derived from $Rpk$ (lower) before  and after meditation.Red and green columns are respectively before and after meditation.Subjects are numbered $1-20$ following name alphabetically }
\label{rpkent}
\end{figure}
\\
The results of $Rpk$ and $ENT$ of each subjects are shown in Fig \ref{rpkent}.The $Rpk$ is increased and $ENT$ goes down for each participant.This indicates improvement of vascular activity of heart in meditative state.Although the percent of change of both parameters varies the trend of change remains same.
\begin{figure}[h!]
\centering
\includegraphics[width=0.9\textwidth]{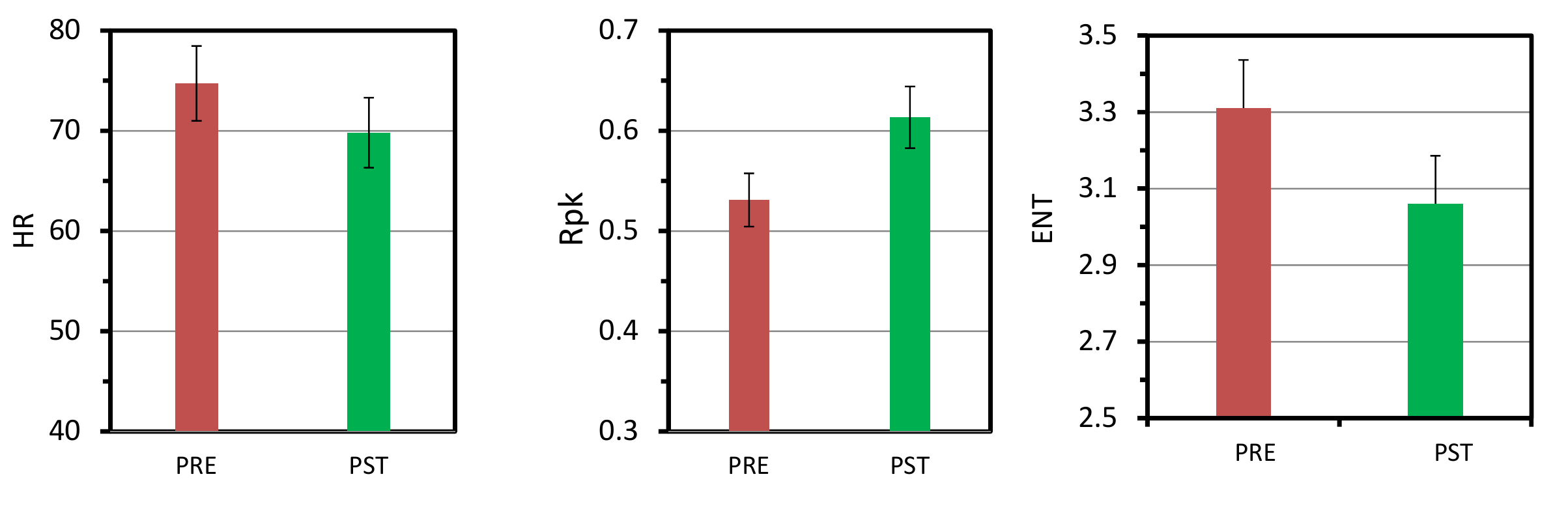}
\caption{ Plot of the heart rate $HR$, the peak value $Rpk$ of $R$ wave and the entropy derived from $Rpk$ data of the group. Red and green columns are respectively Pre- and Post- meditation.}
\label{hrpkent}
\end{figure}
The summary of mean results of heart rate $HR$, the peak value $Rpk$ of $R$ wave and the entropy derived from $Rpk$ data are displayed in Fig.\ref{hrpkent}.The data is statistically significant with $p<0.05$.Due to meditation the heart rate slows down,the ejection goes up.The reduced fluctuation of $R$-peak results lower entropy after meditation.

\begin{figure}[h!]
\centering
\includegraphics[width=1.0\textwidth]{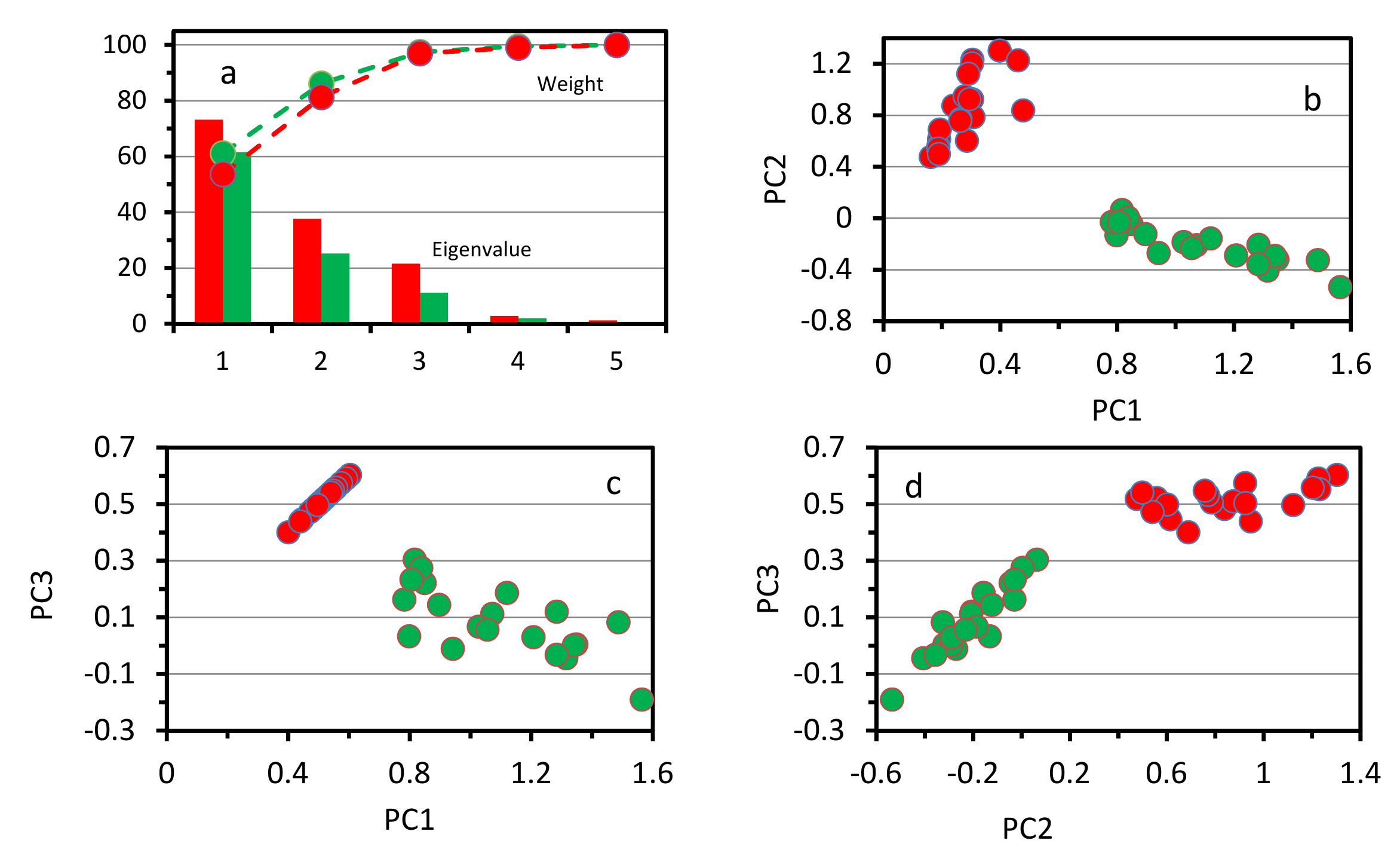}
\caption{The eigenvalues ( bar) and respective weight(line) are in plot (a). The association of first principal components $PC1$ with $PC2$ (b) and $PC3$ (c) and that of $PC2$ with $PC3$ (d) are presented in the plot.The red and green colours refers to pre- and post meditative state.}
\label{pcent}
\end{figure}


The weight of eigenvalue demonstrates that the first three eigenvalues take care of more than $97$ percent of data Fig.(\ref{pcent} a).The first principle component is more important compared to $PC2$ and $PC3$.The association of these $PC$'s resulted from Pre- and Post- meditation are very different Fig.(\ref{pcent} b,c,d).They are well separated in respective phase space and trends are different.

\begin{table}[h!]
\centering
\includegraphics[width=1.0\textwidth]{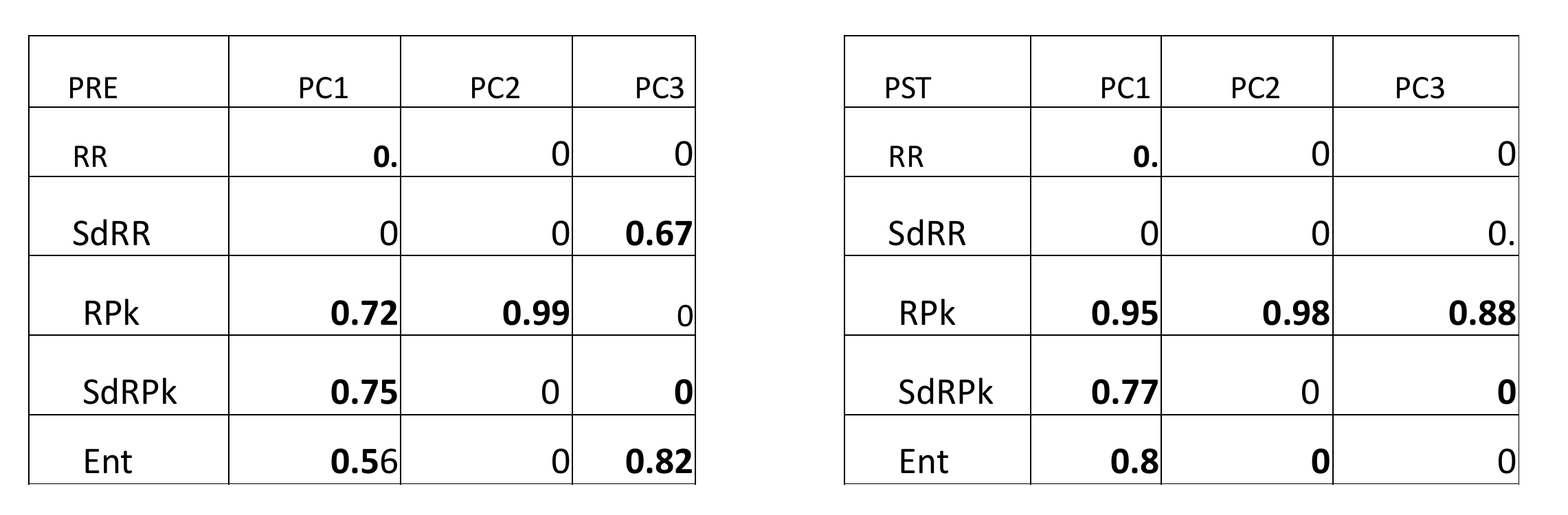}
\caption{The correlation amplitudes among $PC$ and the time-domain variables.}
\label{corpcent}
\end{table}

The correlation between $PC$'s and five parameters are noted in Table.\ref{corpcent}.The non-zero numbers are with $p<0.001$ and zero when $p$ values are much larger than above limit.The analysis shows that the peak of $R$ wave and its fluctuation and the associated entropy may play an important role in assessing the cardiovascular activity.Also it is to be noted that the correlation between these three parameters with $PC1$ changes in Post- meditative state.Although $HR$ is reduced but plays very little role in determining $PC$.The entropy resulting from the fluctuation of heart rate is also considered and change due to meditation are found to be less prominent.\\

\subsection{\textit{Correlation of Fluctuation of Beat Interval and $R$ amplitude}}
The correlation functions ($Corr-RR$) and ($Corr-Rpk$) as a function of lag $m$ for two meditators were obtained using Eq.\ref{cor}.

\begin{figure}[h!]
\centering
\includegraphics[width=1.\linewidth]{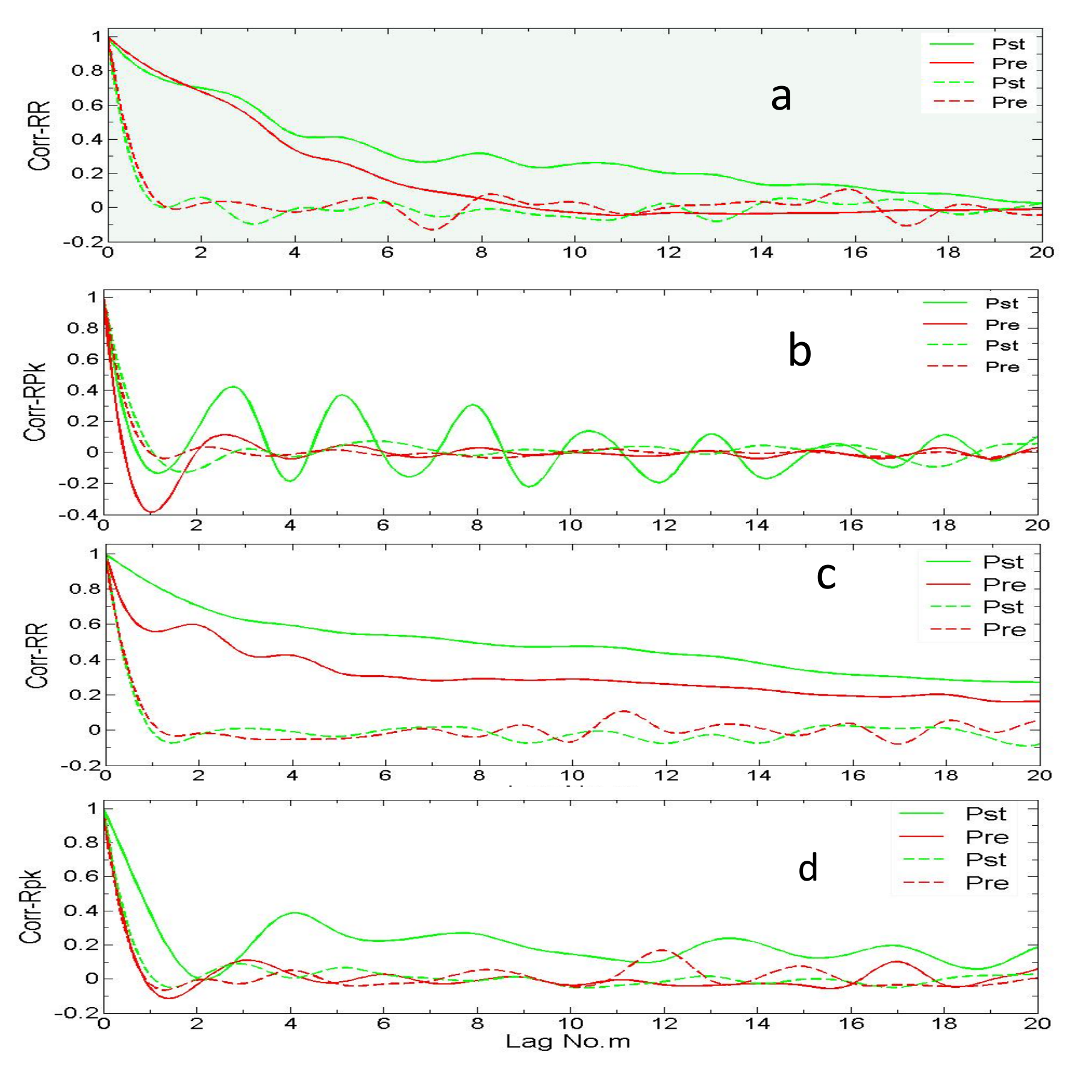}
\caption{Plots are for the correlation functions $Corr-RR$ of $RR$ interval fluctuation and $Corr-Rpk$ of $Rpk$ amplitude fluctuation for different values of lag $m$.Plots (a,b) are for youngest and (c,d) for eldest member in the group.Red and green lines represent data before and after meditation and dotted ones are for shuffled data.}
\label{corr}
\end{figure}

The subjects were chosen considering their experience in the meditative practice.In the group they were respectively oldest,practising for many years and youngest,in the process of mastering the meditation.In order to examine extent of the correlation the data on $RR$ intervals and  $Rpk$ were randomised by shuffling successively $5$ times.Further shuffling did not change the character of the correlation of the shuffled data.
It is clear from Fig.\ref{corr} that both the heart rate and the amplitude of $R$ wave significantly more correlated after(Green line)meditation compared to starting state(Red one).The correlation of the shuffled data are all alike and exhibit random character with near-zero correlation for finite $m$.The small fluctuation around zero is related to error.The correlation of $RR$ fluctuation decays with $m$ with superimposed oscillatory character and remains finite for appreciable number of $RR$.The oscillation aspect of correlation is more prominent for $Rpk$. Though the number of distinct oscillation are found to be different for different pupil but it always exists.

\newpage
\section{\textit{Discussions and Conclusions}}
The influence of (\textit{Dhayana}) a form of meditation regularly practiced by monk of \textit{Ramkrishna Mission} on the HRV are studied using short duration of the ECG. The effects of  meditation on the HRV are assessed from the behavioural changes of the Poincar\'{e} parameters obtained from the Poincar\'{e} plot of the $RR$ interval.The analysis of the Poincar\'{e} parameters and their dynamics clearly reveal beneficial aspect of meditation. The alteration of the cardiovascular regulation are reflected in the difference in distribution of parameters. It has been recognized that any given RR interval influences nearly eight consecutive of heart beats that follow, and this notion triggers the analysis of lagged plot \cite{Thakre},\cite{Lerma}.The lagged Poincar\'{e} plot is found to be more effective in finding out the differences in HRV \cite{Bhaskar},\cite{Sobhendu} on the state of ANS. The parameter $SD1$,the short-time fluctuation. As $SD1$ correlates with short term variability of heart rate and is mainly determined by parasympathetic response,the higher value of $SD1$ after meditation indicates augmentation of parasympathetic response. Autonomic imbalance(increased sympathetic and decreased parasympathetic tone)is known to be associated with increased cardiovascular morbidity and mortality.On the other hand both components of ANS contribute to the
$SD2$ (long term variability) that is more in post meditative state. The Poincar\'{e} parameters are increasing function of lagged value $m$.For large value of $m$, $SD12$ tends to unity as the correlation among beats in time nearly vanishes.Important characteristics of growth of these indices turn out to be slope and curvature for low value of $m$. Out of these six derived quantities the slope and -in particular curvature of $SD12$ in pre - and post- meditation differ significantly. The increase in absolute values of both slope and curvature of the $SD12$ after meditation indicates change of ANS activity that is beneficial to health.The principal component analysis of the variables provides more information embedded in the lagged Poincar\'{e} plots.The first four PC associated with four significant eigenvalues of the covariant matrix of data restore most of the variability of data.The trajectory of PC when plotted with appropriate combination of PC clearly separates Pre- and Post- meditation.It also turns out that the PC which is weighted sum of original variables is dominated by the dynamics of non-linear Poincar\'{e} parameter $SD12$.The correlation of beat interval fluctuation shows that the heart beats are far from random in character.Oscillatory character of the correlation function is more prominent with younger participant. The study with larger set of meditator are needed for understanding of the oscillatory aspect of the correlation.

In conclusion, the comparative strength of the Poincar\'{e} indices and their growth with lag index
and thereafter the principal component analysis might be useful to assess the health benefit of meditation practice. As the analysis of ECG data clearly demonstrate alteration of ANS activity for better cardiovascular activity and 'calmness' of mind,it calls for study with larger meditator group to understand the meditative state. Additionally,similar study of depressed subjects can be utilized to assess extent of improvement with meditation.

\subsection*{Acknowledgement}
The authors is grateful to the disciple of \textit{Ramkrishna Mission,Belur,W.B} for  volunteering.The encouragement, assistance and permitting me to serve Vidyamandir,Belur from Swami Divyananda Maharaj are gratefully acknowledged.I am thankful for consultation and help from cardiologist Dr.Subhra Aditya.
\subsection*{Declaration of Interest}
The author declared no biomedical financial interest or potential conflicts of interest.

\maketitle

\end{document}